\documentclass[final]{aipproc}

\layoutstyle{8x11single}

\SetInternalRegister\hbadness{8000}

\newcommand\doingARLO[2][]{%
  \ifx\mmref\undefined #1\else #2\fi
}

\begin{document}

\title 
       {Neutrinos in Stochastic Media: From Sun to Core-Collapse
         Supernovae} 


\author{A.B. Balantekin}{
  address={Department of Physics, University of Wisconsin, Madison, WI 
53706 USA\thanks{E-mail: baha@nucth.physics.wisc.edu}}
}

\copyrightyear  {2001}

\begin{abstract}
Recent work on neutrino propagation in stochastic media and its
implications for the Sun and core-collapse supernovae are 
reviewed. It is shown that recent results from Sudbury Neutrino
Observatory and SuperKamiokande combined with a best global fit value
of  $\delta m^2= 5\times 10^{-5}$ eV$^2$ and $\tan^2 \theta = 0.3$
rule out solar electron density fluctuations of a few percent or more. 
It is argued that solar neutrino experiments may be able to rule out 
even smaller fluctuations in the near future.   
\end{abstract}

\date{\today}

\maketitle

\indent

Recent observation of the charged-current solar neutrino flux at the
Sudbury Neutrino Observatory \cite{Ahmad:2001an} together with the
measurements of the $\nu_{\odot}$-electron elastic scattering at the
SuperKamiokande detector \cite{Fukuda:2001nj} established that there
are at least two active flavors of neutrinos of solar origin reaching 
Earth. Analyses of all the solar neutrino data updated after the
Sudbury Neutrino Observatory results were announced 
indicate (in two-flavor mixing schemes) a best fit value of $\delta
m^2= 5\times 10^{-5}$ eV$^2$ and $\tan^2 \theta = 0.3$
\cite{Fogli:2001vr,Bahcall:2001zu}. 

In calculating neutrino survival probability in matter one typically
assumes that the electron density of the Sun is a monotonically
decreasing function of the distance from the core and ignores
potentially de-cohering effects \cite{Sawyer:1990tw}. To understand
such effects one possibility is to study parametric changes in the
density \cite{Schaefer:1987fr,Krastev:1989ix}  or the role of matter
currents  \cite{Haxton:1991qb}. Loreti and Balantekin
\cite{Loreti:1994ry} considered neutrino propagation in stochastic
media. They studied the situation where the electron density in the
medium has two components, one average component given by the Standard
Solar Model or Supernova Model, etc., $N_e(r)$, and one fluctuating
component, $N^r_e(r)$. The two-flavor Hamiltonian describing neutrino
propagation in such a medium is given by 
\begin{equation}
\hat H =
\left({{-\delta m^2}\over 4E} \cos 2\theta + {1\over \sqrt{2}}
G_F(N_e(r) + N^r_e(r))\right){\sigma_z} + \left({{\delta m^2}\over 4E}
\sin 2\theta \right) {\sigma_x}.  
\end{equation}
where for consistency one imposes the condition  
\begin{equation}
\langle N^r_e(r)\rangle = 0, 
\end{equation}
and one takes the two-body correlation function to be  
\begin{equation}
\label{bbb}
\langle N^r_e(r)N^r_e(r^{\prime}) \rangle =
{\beta}^2 \ N_e(r) \ N_e(r^{\prime}) \ \exp(-|r-r^{\prime}|/\tau_c).
\end{equation}
In the last equation $\beta$ denotes the amplitude of the
fluctuations. In the calculations of the Wisconsin group the
fluctuations are typically taken to be subject to colored noise,
i.e. higher order correlations 
\begin{equation}
f_{12 \cdots }=\langle N^r_e(r_1)N^r_e(r_2) \cdots
\rangle
\end{equation}
are taken to be 
\begin{equation}
 f_{1234}= f_{12}f_{34} + f_{13}f_{24} +
f_{14}f_{23},
\end{equation} 
and so on.

\begin{figure}
\caption{Mean electron neutrino survival probability plus minus
$\sigma$ in the Sun with fluctuations. The average electron density is
given by the Standard Solar Model of Bahcall and collaborators
\cite{Bahcall:2000nu} and  
$\tan^2 2 \theta=0.3$. The amount of the fluctuation (the value of
$\beta$ in Eq. (\ref{bbb})) used in
calculating each panel is indicated.
For completeness the effects of an unrealistic fluctuation of $\beta$
= 50 \% are also shown. 
}
\includegraphics[height=.5\textheight]{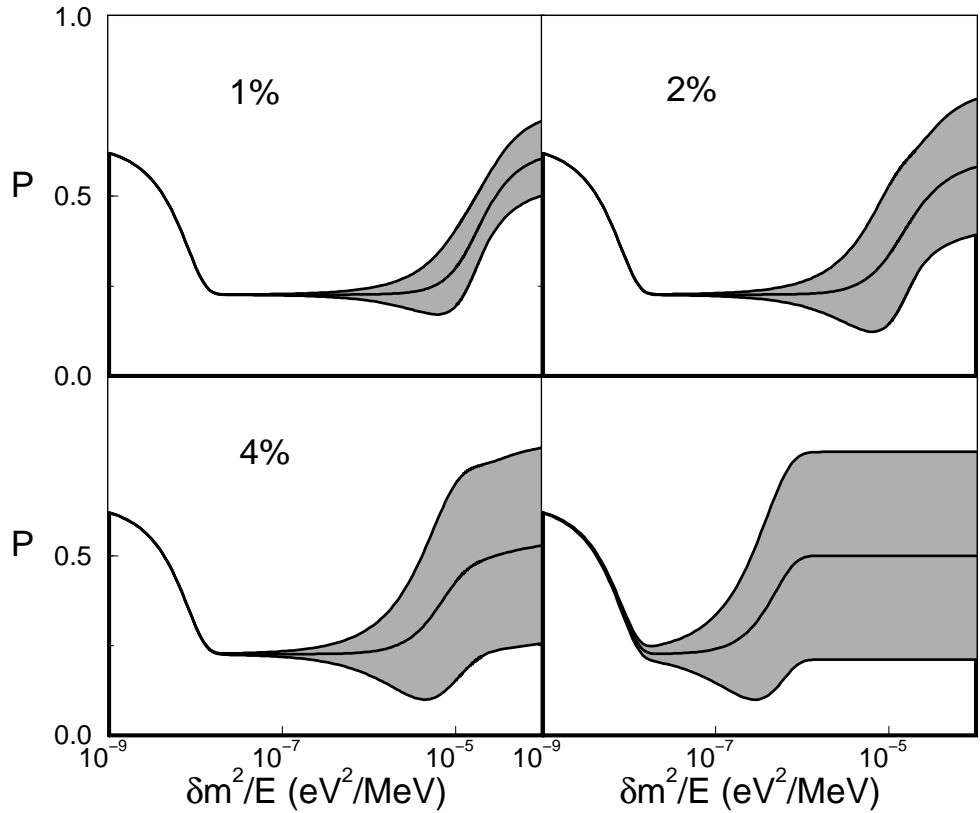}
\end{figure}

Using the
formalism sketched above, it is possible to calculate not only the
mean survival probability, but also the variance, $\sigma$,  of the
fluctuations to get a feeling for the distribution of the survival
probabilities \cite{Balantekin:1996pp} as illustrated in Figure 1.
One notes that for very large
fluctuations complete flavor de-polarization should be achieved,
i.e. the neutrino survival probability is 0.5, the same as the vacuum
oscillation probability with two flavors for long distances. To
illustrate this behavior results from the physically unrealistic
case of $\beta$ = 50\% fluctuations are shown as well in the Figure 1
(the 
lower right-hand panel). One observes that for large values of $\delta
m^2$ the average survival probability is indeed $0.5$ with a spread
between a survival probability of $0$ and $1$ ($0.5 \pm 0.5$).  
Detailed investigations indicate that the effect of the
fluctuations is largest when the neutrino propagation in their absence
is adiabatic \cite{Loreti:1995ae,Balantekin:1996pp}. Note that the
fluctuation analysis presented here is for two active flavors. This is
exact only in the limit $U_{e3}=0$ \cite{Balantekin:1999dx} when
the solar electron neutrinos mix with an appropriate linear
combination of the $\nu_{\mu}$'s and $\nu_{\tau}$'s determined
by the atmospheric neutrino measurements. When $U_{e3} \neq 0$ a full 
three-flavor analysis is needed except for very special
circumstances. However, since global analysis indicates that  $U_{e3}
\sim 0$ \cite{Fogli:2001vr} two-neutrino analysis may be rather
accurate in the absence of sterile neutrinos. 

In these calculations the correlation length $\tau_c$ is taken to be
very small, of the order of 10 km., to be consistent with the
helioseismic observations of the sound speed
\cite{Bahcall:2000nu,Bahcall:1997fu}. In the 
opposite limit of very large correlation lengths are very interesting
result is obtained \cite{Loreti:1995ae}, namely the averaged density
matrix is given as an integral 
\begin{equation}
\lim_{\tau_c\to\infty}\langle\hat \rho(r)\rangle =
{1\over{\sqrt{2\pi \beta^2}}} \int_{-\infty}^{\infty} dx
\exp[{-x^2/(2\beta^2)}]
\hat \rho(r,x),
\end{equation}
reminiscent of the channel-coupling problem in nuclear
physics \cite{Balantekin:1998yh}. Even though this limit is not
appropriate to 
the solar fluctuations it may be applicable to a number of other
astrophysical situations where neutrinos play a role. Similar
conclusions were reached by other authors 
\cite{Burgess:1997mz,Nunokawa:1997dp,Bamert:1998jj,Nunokawa:1997tq,%
Sahu:1998jh,Bykov:1998gv,Valle:1999gj,Bykov:2000ze,Reggiani:1998qy,%
Pantaleone:1998xi,Torrente-Lujan:1998mc,Torrente-Lujan:1998pf,%
Prakash:2001rx,Bell:2000ir}. Recent reviews were given in Refs. 
\cite{Balantekin:1998yb,Valle:2001kt}. 

\begin{figure}
\caption{Solar Neutrino flux. The error range in the prediction of the
  Standard Solar Model is indicated by the black band. The spread of
  the electron neutrino flux due to $\beta$ = 1\% fluctuations is
  indicated by the shaded area (the curve in the middle of the shaded
  area is the prediction for the average survival probability in the
  presence  of the fluctuations with
  $\delta m^2= 5\times 10^{-5}$ eV$^2$ and $\tan^2 \theta = 0.3$. The
  total energy-integrated solar electron neutrino flux observed in
  Sudbury Neutrino Observatory (the horizontal dashed line), the
  corresponding experimental 
  error (two horizontal lines bordering the dashed line), as well as
  the spread of the total energy-integrated solar neutrino flux of all
  active flavors evaluated using the Sudbury Neutrino Observatory and
  SuperKamiokande data 
  (three very closely spaced horizontal lines at the top of the
  figure) are also shown.}
\includegraphics[height=.4\textheight]{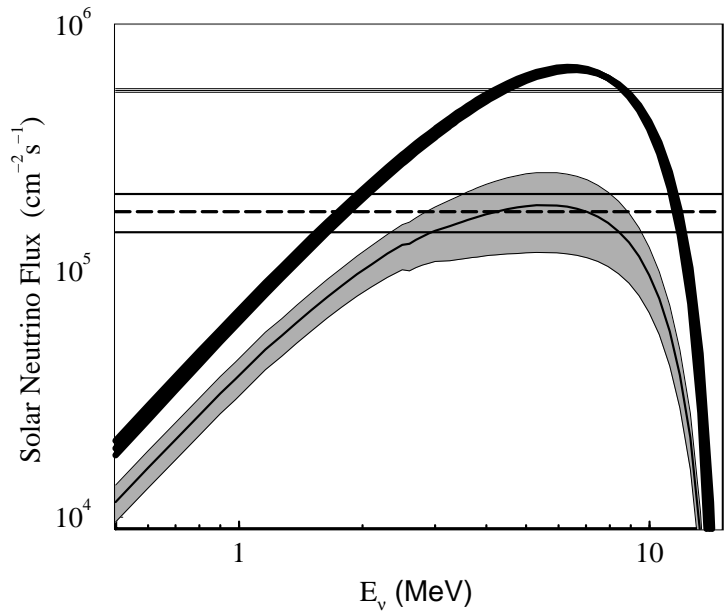}
\end{figure}

Recent Sudbury Neutrino Observatory measurements \cite{Ahmad:2001an}
give the electron neutrino 
component of the solar $^8$B flux to be $1.75 \pm 0.15 \times 10^6$
cm$^{-2}$ s$^{-1}$. This result combined with the SuperKamiokande
measurements \cite{Fukuda:2001nj} give the flux of total active flavor
component of the $^8$B neutrinos to be $5.44 \pm 0.99 \times 10^6$
cm$^{-2}$ s$^{-1}$ in good agreement with the Standard Solar Model
prediction of $5.05 \pm 0.80 \times 10^6$ cm$^{-2}$ s$^{-1}$
\cite{Bahcall:2000nu}. (The theoretical error is large due to the
large experimental uncertainty in the nuclear reactions forming $^8$B
\cite{Adelberger:1998qm}). These data are shown in Figure 2. The best
fit to all the solar neutrino data is with $\delta
m^2= 5\times 10^{-5}$ eV$^2$ and $\tan^2 \theta = 0.3$. The mean solar
electron neutrino flux in the presence  of the fluctuations as well as
its variance are also shown. 

Figure 2 suggests that 
the Sudbury Neutrino Observatory charged-current measurements safely
rule out solar electron density fluctuations of a few percent or
more as such fluctuations would give rise to a much larger spread of
the solar electron neutrino flux than that was observed (Strictly
speaking the time-scale associated with the 
fluctuations should be incorporated in this argument; in the present
discussion this time-scale is assumed to be much shorter than the
total data-taking duration). A careful analysis of the time-structure
of the future charged-current data at Sudbury Neutrino Observatory
and the measurements of the $^7$Be-line at BOREXINO and KamLAND   
may be able to rule out fluctuations of much smaller amplitudes. Note
that it is essential to take the variance into account in such an
analysis; the mean survival probability in the presence of
fluctuations is too close to the survival probability in the absence
of fluctuations when $\beta=1$\%.    

The solar electron density fluctuations cause 
not only fluctuations in the solar $\nu_e$ flux, but, because of the
unitarity, compensating fluctuations in the solar $\nu_{\mu}$ and
$\nu_{\tau}$ flux. Consequently it is harder to perform such an
analysis with $\nu_{\odot}$-electron elastic scattering since elastic
scattering is sensitive to all flavors (albeit at a reduced rate for
muon and tau neutrinos). A more detailed analysis of the possible
signatures of fluctuations in the solar neutrino data will be
published elsewhere \cite{baha}. 

Neutrino propagation in the presence of fluctuations was also
investigated for the neutrino convection in a core-collapse
supernova where the adiabaticity condition is well satisfied
\cite{Loreti:1995ae}. In core-collapse supernovae neutrinos not only 
may play a very important role in shock re-heating
\cite{Prakash:2001rx}, but may also help solving a number of problems
arising when such supernovae are considered as possible sites for 
r-process nucleosynthesis
\cite{McLaughlin:1999pd,Caldwell:1999zk}. Fluctuations are likely to
be 
important in assessing the role of neutrinos in such phenomena. 

\begin{theacknowledgments}
This work was supported in part by the U.S. National  Science
Foundation Grant No.\ PHY-0070161  at the University of Wisconsin, and
in part by the University  of Wisconsin Research Committee with funds
granted by the Wisconsin Alumni Research Foundation.
\end{theacknowledgments}


\doingARLO[\bibliographystyle{aipproc}]
          {\ifthenelse{\equal{\AIPcitestyleselect}{num}}
             {\bibliographystyle{arlonum}}
             {\bibliographystyle{arlobib}}
          }

\bibliography{fluc}

\end{document}